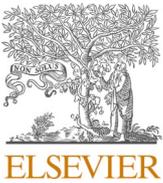
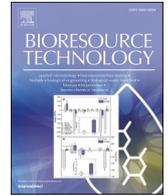

# Effect of turbulent diffusion in modeling anaerobic digestion

Jeremy Z. Yan, Prashant Kumar, Wolfgang Rauch[*]

*Environmental Engineering Department, University of Innsbruck, Technikerstraße 13, Room 312, 6020 Innsbruck, Austria*

HIGHLIGHTS

- Turbulent diffusion models are implemented for both thermal and chemical.
- The implementation is validated and shows impact on anaerobic digestion process.
- Thermal turbulent diffusion has more significant impact on anaerobic digestion.

GRAPHICAL ABSTRACT

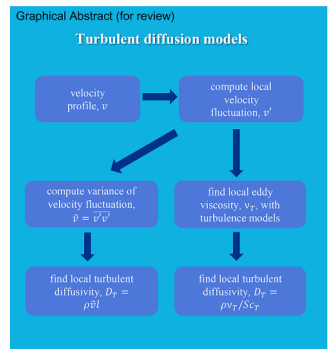
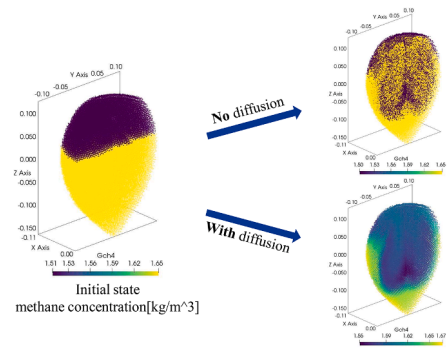

ABSTRACT

In this study, the impact of turbulent diffusion on mixing of biochemical reaction models is explored by implementing and validating different models. An original codebase called CHAD (Coupled Hydrodynamics and Anaerobic Digestion) is extended to incorporate turbulent diffusion and validate it against results from OpenFOAM with 2D Rayleigh-Taylor Instability and lid-driven cavity simulations. The models are then tested for the applications with Anaerobic Digestion - a widely used wastewater treatment method. The findings demonstrate that the implemented models accurately capture turbulent diffusion when provided with an accurate flow field. Specifically, a minor effect of chemical turbulent diffusion on biochemical reactions within the anaerobic digestion tank is observed, while thermal turbulent diffusion significantly influences mixing. By successfully implementing turbulent diffusion models in CHAD, its capabilities for more accurate anaerobic digestion simulations are enhanced, aiding in optimizing the design and operation of anaerobic digestion reactors in real-world wastewater treatment applications.

## 1. Introduction

The understanding of chemical and biochemical processes is of paramount importance in modern process industries. Numerical models play a crucial role in providing researchers and engineers with a cost-effective and simplified alternative to experiments. However, the biochemical reaction models typically used in these industries, for example the Anaerobic Digestion Model No.1 (ADMno1) (Batstone et al., 2002), assumes the complete homogeneity of the reactor, akin to an idealized Continuous-Stirred Tank Reactor (CSTR). In real-life applications, biochemical reactors often suffer from inadequate mixing, resulting in reduced efficiency and production rates. Harris et al. (1996) has shown the positive impact of coupling Eulerian Computational Fluid Dynamics (CFD) with chemical reaction models on the improvements in accuracy. Previous Studies (Bałdyga and Pohorecki, 1995; Dapelo and Bridgeman, 2018) emphasized the significance of incorporating CFD in chemical reaction simulations and the effect of mixing. Coughtrie et al. (2013) emphasized the importance of turbulent modelling on anaerobic gas-lift digester.

**Table 1**
Parameters for the RTI simulation.

| Parameter Name | Value |
| --- | --- |
| Height (transversal) | 0.12 m |
| Width (longitudinal) | 0.04 m |
| Density (heavy) | 1500 kg/m³ |
| Density (light) | 1000 kg/m³ |
| Temperature (warm) | 300 K |
| Temperature (cold) | 298 K |
| Thermal conductivity | 1E−8 W/(m*K) |
| Specific heat capacity | 4180 J/(kg*K) |
| Turbulent Prandtl number | 0.01 |

**Table 2**
Parameters for the lid-driven cavity simulation.

| Parameter Name | Value |
| --- | --- |
| Height (transversal) | 1.0 m |
| Width (longitudinal) | 1.0 m |
| Density | 1000 kg/m³ |
| Kinematic viscosity | 1E−4 m²/s |
| Driving velocity | 1 m/s |
| Schmidt number | 1.0E20 |
| Turbulent Schmidt number | 0.1 |
| Scalar (high concentration) | 1.1E−3 kg/m³ |
| Scalar (low concentration) | 1.0E−3 kg/m³ |

**Table 3**
Parameters for the lab-scale tank simulation.

| Parameter Name | Value |
| --- | --- |
| Volume of tank | 8E−3 m³ |
| Mixer velocity | 12 rpm |
| Sludge temperature (warm) | 309.15 K |
| Sludge temperature (cold) | 308.15 K |
| wall temperature (mixer) | 310.15 K |
| wall temperature (wall) | 303.15 K |
| Volume of particle | 6.4E−8 m³ |
| Number of Particles | 128,726 |
| Simulation Duration | 200 s |

**Table 4**
Concentration of inlet for lab-scale tank simulation.

| Variable | Value | Unit |
| --- | --- | --- |
| $S_{su,in}$ | 0 | kg·COD·m⁻³ |
| $S_{aa,in}$ | 0.044 | kg·COD·m⁻³ |
| $S_{fa,in}$ | 0 | kg·COD·m⁻³ |
| $S_{va,in}$ | 0 | kg·COD·m⁻³ |
| $S_{bu,in}$ | 0 | kg·COD·m⁻³ |
| $S_{pro,in}$ | 0 | kg·COD·m⁻³ |
| $S_{ac,in}$ | 0 | kg·COD·m⁻³ |
| $S_{h2,in}$ | 0 | kg·COD·m⁻³ |
| $S_{ch4,in}$ | 0 | kg·COD·m⁻³ |
| $S_{IC,in}$ | 0.008 | kmole C·m⁻³ |
| $S_{IN,in}$ | 0.002 | kmole N·m⁻³ |
| $S_{I,in}$ | 0.028 | kg·COD·m⁻³ |
| $X_{c,in}$ | 0 | kg·COD·m⁻³ |
| $X_{ch,in}$ | 3.72 | kg·COD·m⁻³ |
| $X_{pr,in}$ | 16.9 | kg·COD·m⁻³ |
| $X_{li,in}$ | 8.05 | kg·COD·m⁻³ |
| $X_{su,in}$ | 0 | kg·COD·m⁻³ |
| $X_{aa,in}$ | 0 | kg·COD·m⁻³ |
| $X_{fa,in}$ | 0 | kg·COD·m⁻³ |
| $X_{c4,in}$ | 0 | kg·COD·m⁻³ |
| $X_{pro,in}$ | 0 | kg·COD·m⁻³ |
| $X_{ac,in}$ | 0 | kg·COD·m⁻³ |
| $X_{h2,in}$ | 0 | kg·COD·m⁻³ |
| $X_{I,in}$ | 17.0 | kg·COD·m⁻³ |
| $S_{cat,in}$ | 0 | kmole·m⁻³ |
| $S_{an,in}$ | 0.0052 | kmole·m⁻³ |
| $q_{in}$ | 178 | m³·d⁻¹ |
| $T_{op}$ | 35 | °C |

In this study, the aforementioned Anaerobic Digestion (AD) reaction and ADMno1 are selected to be the study case since they are well researched and validated (Batstone et al., 2002; Rosén and Jeppsson, 2006). The significance of mixing in AD reaction tanks is demonstrated by the work of Lindmark et al. (2014). To address the limitations associated with the assumption of homogeneity, several attempts have been made to couple CFD with ADMno1. Rezavand et al. (2019) integrated the Lagrangian CFD method, Smooth Particle Hydrodynamics (SPH) (Monaghan, 1988), with ADMno1 to study a two-dimensional AD reaction tank with passive mixing. Dabiri et al. (2023) combined ADMno1 with the Eulerian CFD method to investigate the mixing effects of a recirculation mixing device. Similarly, Tobo et al. (2020) employed a compartmental model approach to link Eulerian CFD to ADMno1 and arrived at the same conclusion that assuming homogeneity within the AD reaction tank leads to notable inaccuracies.

During the initial stages of this project (Kumar et al., 2023)s, an original codebase named CHAD (Coupled Hydrodynamics and Anaerobic Digestion) was developed to incorporate additional details such as thermal and chemical mixing effects by combining data from SPH simulations with ADMno1. The studied case involves a three-dimensional AD reaction tank with an active mixing device (helical rotor), based on a real-life lab-scale model. While CHAD shows significant improvements compared to conventional ADMno1, the current model for mixing effects assumes a constant diffusion coefficient across the computational domain for both thermal and chemical dynamics. This approach is suboptimal for simulating devices where fluids exhibit nontrivial velocity and thermal gradients. The diffusion of scalars, such as chemical concentrations and temperature, is primarily driven by molecular and turbulent diffusion. Molecular diffusion arises from the random thermal motion of molecules or scalar gradients and is known as Brownian or Fickian diffusion, respectively. The diffusion coefficients for molecular diffusion of different substances are at the scale of 1E−9 $m^2/s$ (Holz et al., 2000), negligible compared with the diffusion caused by the velocity gradient. In anaerobic digestion tanks with passive or active mixing mechanisms, turbulence often plays a dominant role in mixing. Therefore, dynamically calculating the diffusion coefficient based on local turbulence characteristics becomes crucial.

The effect of turbulence on scalar mixing was first studied by Taylor (1922), where he statistically analyzed the essential properties of turbulent motion that transport and diffuse scalars. He found that the statistical characteristics of fluid particles are characterized by the variance of the background velocity field and the correlation coefficient. Roberts and Webster (2003) proposed a method that utilizes the variance of velocity fluctuations and a defined characteristic length scale to determine turbulent diffusion coefficients. Greif et al. (2009), on the other hand, introduced a method that uses velocity dispersion and characteristic length scale to calculate the effective diffusion coefficient, which combines turbulent and laminar diffusion coefficients. Another classic approach for modeling turbulent diffusion coefficients is the application of the Gradient-Diffusion Hypothesis (GDH) (Combest et al., 2011), where it assumes that the turbulent diffusion coefficient is linearly proportional to the turbulent viscosity (eddy viscosity).

This project aims to investigate the modelling methods of turbulent diffusion in the particle-based computational fluid dynamics schemes and the influence of turbulent diffusion on the mixing effect in chemical reactions. It is projected that turbulent diffusion plays a significant role in flows with considerable velocity gradients, and its impact cannot be overlooked. By implementing and validating different turbulent diffusion models in CHAD, it is aimed to enhance the understanding of turbulent diffusion modelling and the turbulent mixing effects in chemical reaction simulations, thereby contributing to a more comprehensive understanding of the topic.

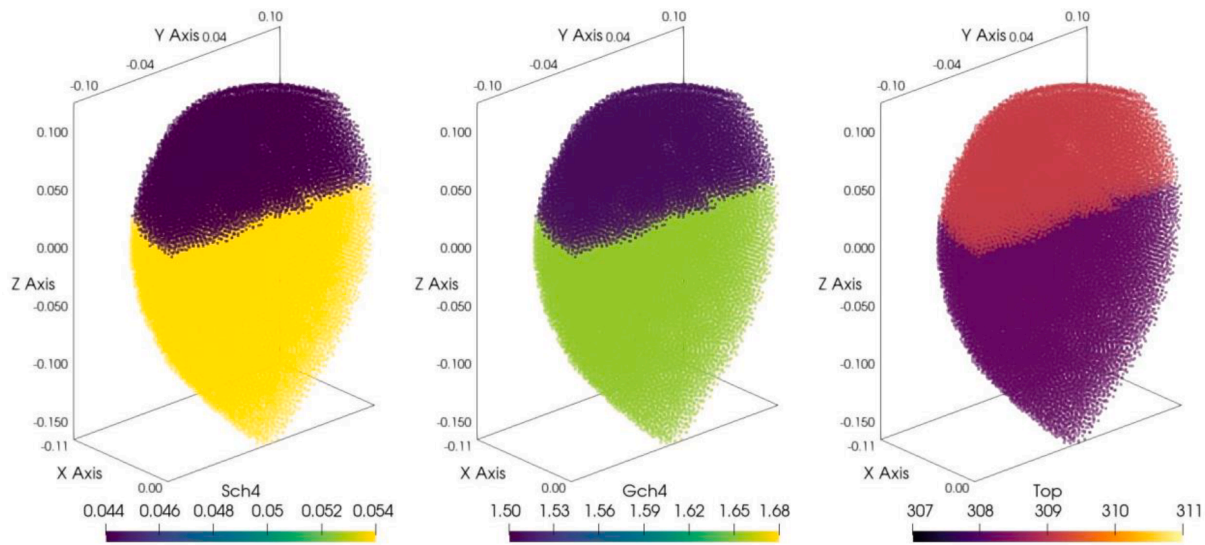

**Fig. 1.a.** The initial condition of the lab-scale reactor. Soluble methane, gaseous methane and operating temperature fields from left to right.

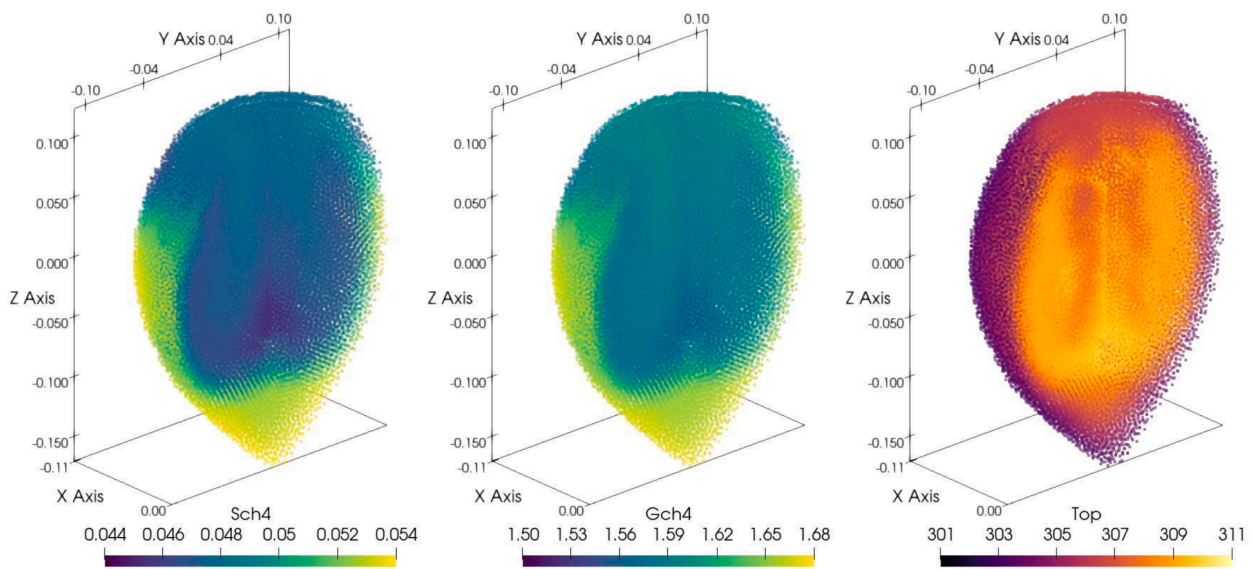

**Fig. 1.b.** The lab-scale reactor with turbulent diffusion (Roberts and Webster, 2003) implemented after 200 s. Soluble methane, gaseous methane and operating temperature fields from left to right.

## 2. Methodology

### 2.1. Anaerobic Digestion Model No.1

Anaerobic digestion is a biological process that converts organic waste into biogas and fertilizer. The ADMno1 model was proposed by Batstone et al. (2002) and has since become one of the most widely used mathematical models in the industry and academia for AD process. The model takes in factors like inhibition, pH value and temperature, to calculate the reaction rates for the 27 main components and several other intermediate components. For components that could be both gaseous and soluble, carbon dioxide and methane for example, the model also includes the gas transfer process, which takes pressure and temperature as parameters. In this work, the implementation of ADMno1 is based on the work of Rosen and Jeppsson (2006). Since each particle is modeled as a CSTR, the following mass balance equation is established Eq. (1),

$$\frac{dVS_i}{dt} = q_{in}S_{in,i} - q_{out}S_i + V\sum_j \dot{\rho}_j v_{i,j} \qquad (1)$$

where $V$ is the volume of the CSTR, or the local particle, and $S_i$ and $S_{in,i}$ are the concentration of component $i$ in the tank and of inlet respectively, and $q_{in}$ and $q_{out}$ are the flow rates of inlet and outlet. The index $j$ indicates a certain biochemical process and $\dot{\rho}_j$ and $v_{i,j}$ are the kinetic rate (the term "reaction rate" is also used for $\dot{\rho}_j$ interchangeably) and biochemical rate coefficient for component $i$ in that process $j$, respectively. $v_{i,j}$ is calculated as the stoichiometry table stated in the work of Rosen and Jeppsson (2006). And the kinetic rate $\dot{\rho}_j$ is calculated using reaction rate constants $k_m$, and inhibition parameters such as $K_S$, $K_I$ and inhibition factors $I$. In the current implementation of ADMno1 in this work, the pre-defined parameters are set in the beginning of the simulation, based on different operation conditions, such as mesophilic and thermophilic temperature schemes (Hansruedi Siegrist, 2002; Rosén and Jeppsson, 2006). The constants and parameters used as benchmark by Rosén and Jeppsson (2006) are listed in Table 4 and 5 (Batstone et al., 2002).

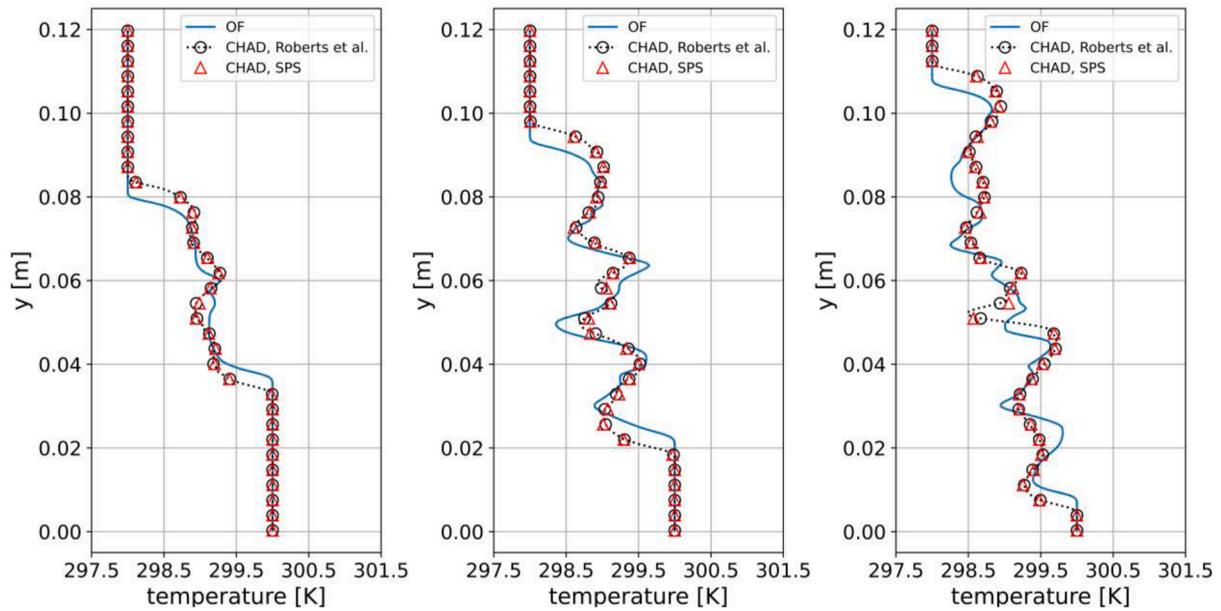

**Fig. 2.a.** The average temperature profile along y-axis for OF and CHAD simulation, with and without thermal diffusion. t = 0.4s, t = 0.6s and t = 0.8s from left to right.

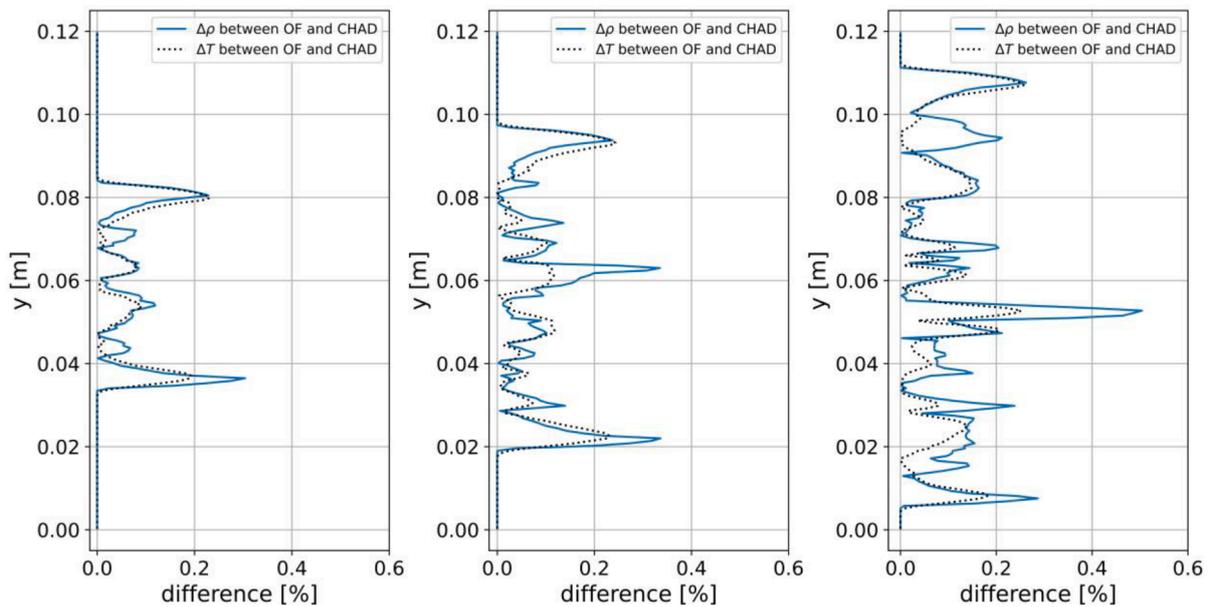

**Fig. 2.b.** The difference of results between OF and CHAD simulation for average temperature and density along y-axis, with and without thermal diffusion. t = 0.4s, t = 0.6s and t = 0.8s from left to right.

The typical form of the equation for calculating the reaction rate $\dot{\rho}$ is shown as Eq. (2) below. Note that for different process $j$ and different component $i$, the form of the equation might be different. Eq. (2) is shown as an example, in which, $k_m$ is the reaction constant, $S_i$ is the concentration of some soluble component $i$, and $X_i$ is the concentration of the biomass that degrades the component $i$.

$$\dot{\rho}_j = k_m \frac{S_i}{K_{S,j} + S_i} X_i I \quad (2)$$

The inhibition factors $I$ take the form of pH-dependent inhibition $I_{pH}$ and non-pH-dependent inhibition $I_m$, where $m$ indicates the component that is being inhibited. The equations for these 2 factors are shown by Eqs. (3) and (5) below, where $K_{I,m}$ is the inhibition parameter of component $m$, $pH_{LL}$ and $pH_{UL}$ are the lower and upper limit of pH, and $n$ is empirical. Such function for calculating $I_{pH}$ is called Hill function,

$$I_m = \frac{S_m}{S_m + K_{I,m}} \quad (3)$$

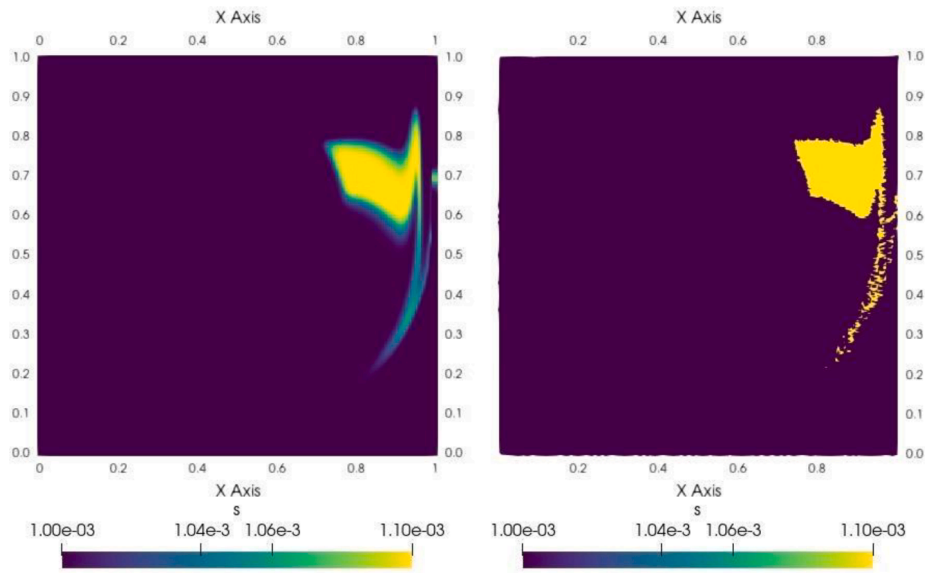

**Fig. 3.a.** The scalar profile, without diffusion for OF (left) and CHAD (right), t = 2s.

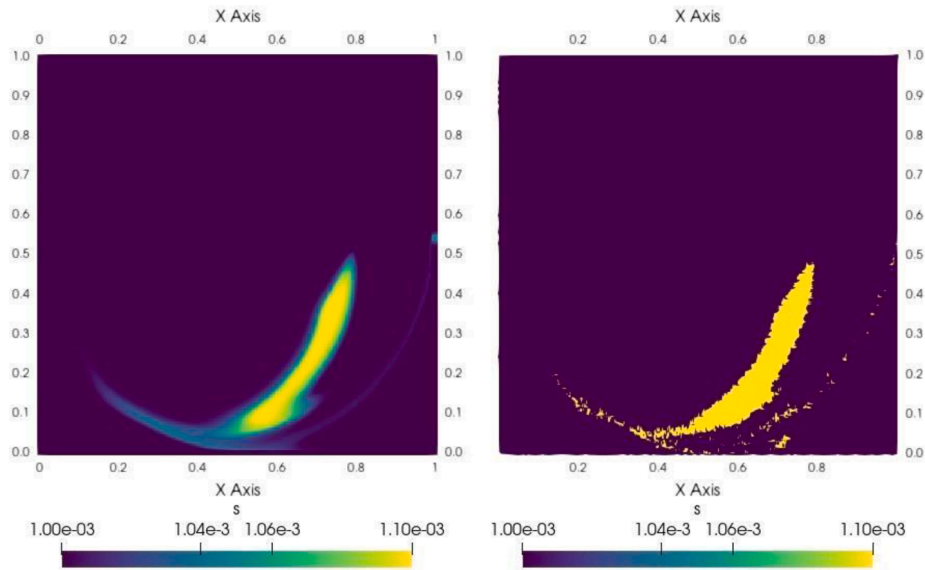

**Fig. 3.b.** The scalar profile, without diffusion for OF (left) and CHAD (right), t = 4s.

$$K_{pH} = \frac{pH_{LL} + pH_{UL}}{2} \tag{4}$$

$$I_{pH} = \frac{pH^n}{K_{pH}^n + pH^n} \tag{5}$$

The pH value is a critical parameter in anaerobic digestion, as it affects the activity of microorganisms and the stability of the process. However, accurately modeling the dynamics of pH using differential equations (DEs) can be challenging due to the fast acid-base reaction rates involved. To accurately and stably calculate the pH value with a reasonable time stepping size, a set of differential–algebraic equations (DAEs) are solved iteratively to find the concentration of proton ($S_{H^+}$) and soluble hydrogen ($S_{h2}$). These equations are based on the principles of mass conservation, charge balance, and chemical equilibrium. In this work, the Newton-Raphson method is used to solve the DAEs for obtaining the values of $S_{H^+}$ and $S_{h2}$. The DAEs are shown in Eqs. (6) and (7). (Rosén and Jeppsson, 2006)

$$E(S_{H^+}) = S_{cat^+} + S_{nh^+} + S_{H^+} - S_{hco3^-} - \frac{S_{ac^-}}{64} - \frac{S_{pr^-}}{112} - \frac{S_{bu^-}}{160} - \frac{S_{va^-}}{208} - \frac{K_W}{S_{H^+}} - S_{an^-} \tag{6}$$

$$E(S_{h2}) = \frac{q_{in}}{V}\left(S_{h2,in} - S_{h2}\right) + \rho_{S_{h2}} - \rho_{G_{h2}} \tag{7}$$

where $\rho_{S_{h2}}$ is the rate of uptake of soluble hydrogen and $\rho_{G_{h2}}$ is the production rate of the gas phase hydrogen.

### 2.2. Smooth particle hydrodynamics

Smooth Particle Hydrodynamics (SPH) is a widely used computational method in fluid mechanics and solid mechanics. It has several advantages over traditional numerical methods like finite volume method (FVM) or finite difference method (FDM). One of the major benefits of SPH is its ability to simulate complex and large-scale problems with moving boundaries, free surfaces, and discontinuities. It is also advantageous that the mass advection is directly modeled by the movement of the particles, therefore avoiding the numerical errors introduced from the discretization schemes. The SPH method is Lagrangian in nature, meaning that the particles follow the motion of the fluid or solid they represent. This allows for the simulation of large deformations and movements, such as those seen in fluid dynamics or solid mechanics problems. Additionally, the Lagrangian approach makes it easy to track the motion of specific particles and to study their individual behavior over time. In SPH, the continuum is discretized into a set of particles with a mass, position, velocity, and other relevant physical properties. The physical properties of the said continuum is governed by the Eq. (9), where $a$ indicates the local particle, $b$ indicates the neighboring particles, $F$ is the studied physical property, and $W(r_a - r_b, h)$ is the smooth kernel function and it has 2 arguments, the distance between local particle $a$ and its neighboring particle $b$, $(r_a - r_b)$, and the radius of the smooth kernel $h$, which is also referred to as the smoothing length. It can be written as $W_{ab}$ and with the same principle, $(r_a - r_b)$ is written as $r_{ab}$. While there are different options for the smooth kernel function the Wendland Quintic (Wendland, 1995) function is applied in the following.

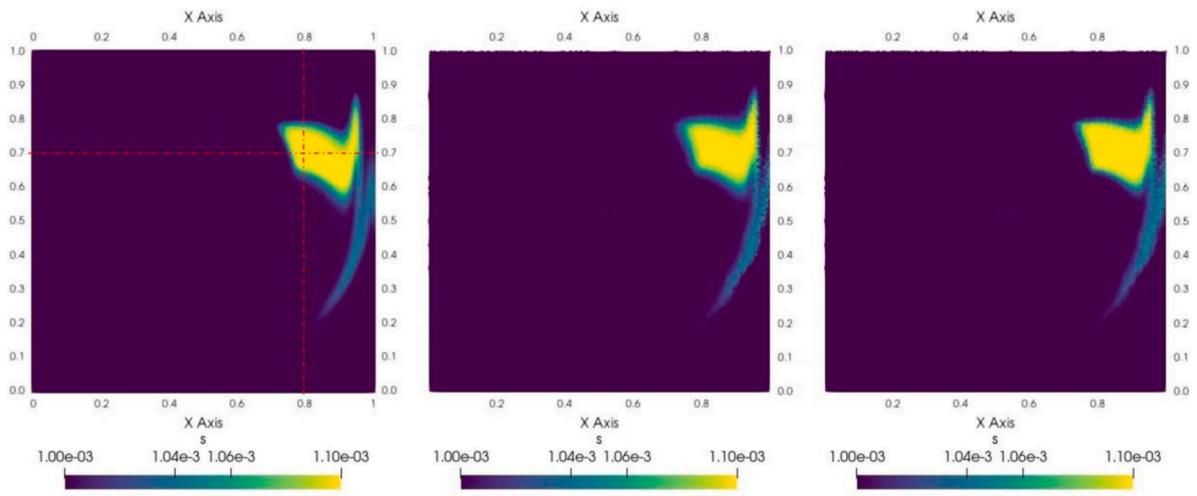

**Fig. 4.a.** The scalar profile with turbulent diffusion, for OF (left), CHAD with Roberts et al. (mid) and CHAD with SPS (right), t = 2s.

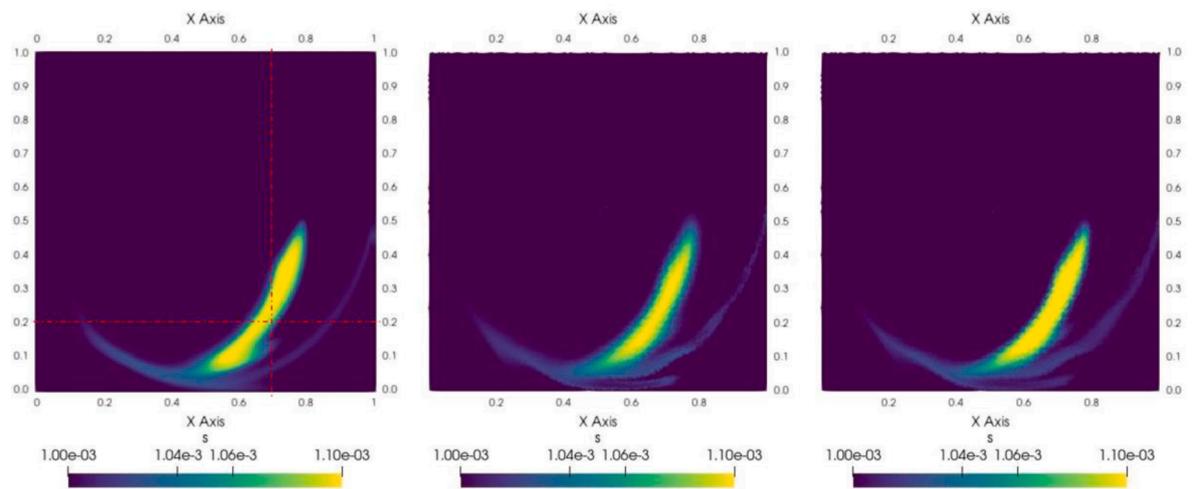

**Fig. 4.b.** The scalar profile with turbulent diffusion, for OF (left), CHAD with Roberts et al. (mid) and CHAD with SPS (right), t = 4s.

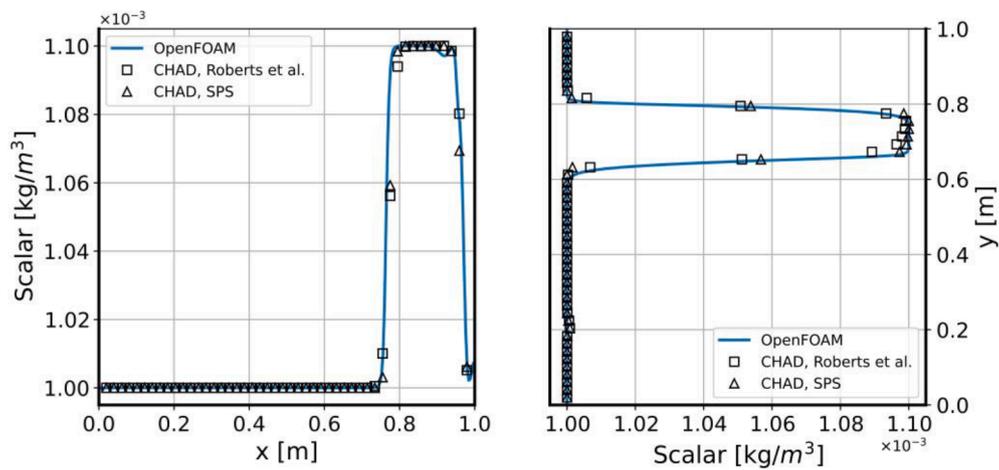

**Fig. 5.a.** The comparison of scalar profile between OF and CHAD with different turbulent diffusion models along x-axis (y = 0.7 m) (left) and along y-axis (x = 0.8 m) (right), t = 2 s.

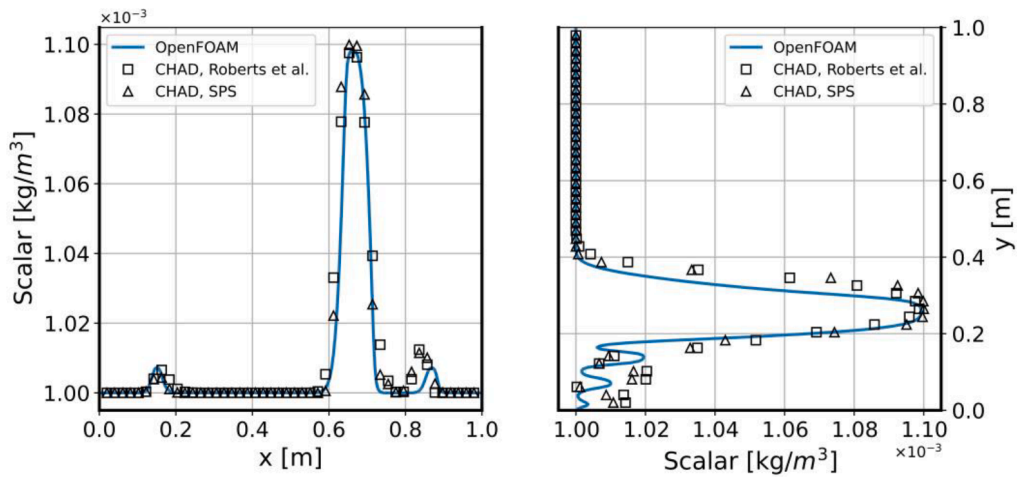

**Fig. 5.b.** The comparison of scalar profile between OF and CHAD with different turbulent diffusion models along x-axis (y = 0.2 m) (left) and along y-axis (x = 0.7 m) (right), t = 2 s.

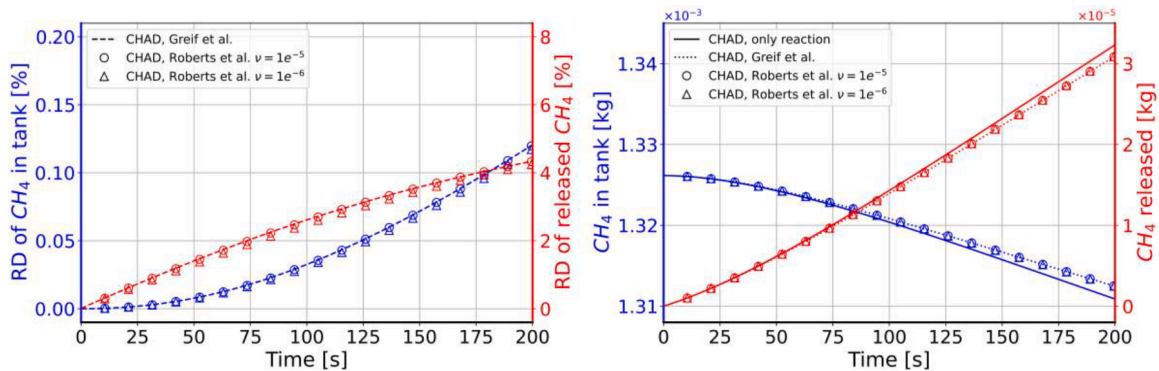

**Fig. 6.a.** The comparison of methane production of lab-scale reactor over 200 s between Greif et al, and Roberts et al. with different kinematic viscosity. Relative difference shown on the left and absolute mass shown in the right. All simulations are set to exclude the thermal turbulent diffusion.

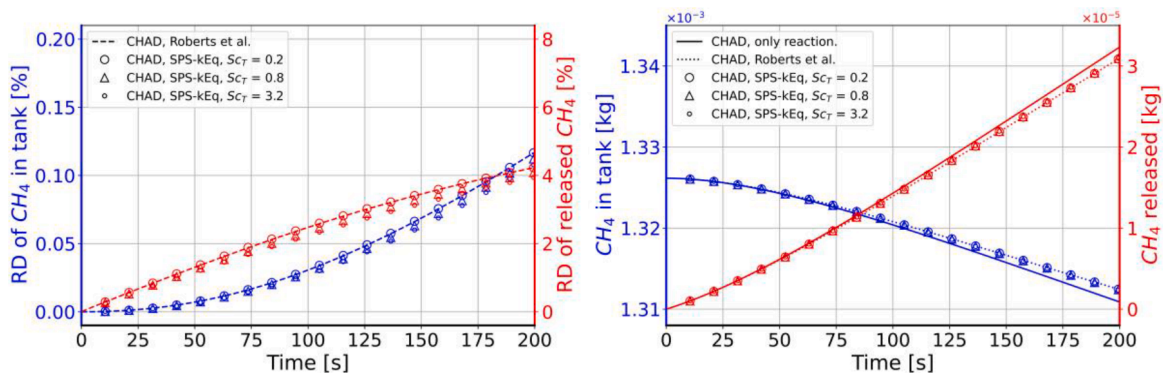

**Fig. 6.b.** The comparison of methane production of lab-scale reactor over 200 s with different turbulent Schmidt number for different models. Relative difference shown on the left and absolute mass shown in the right. All simulations are set to exclude the thermal turbulent diffusion. The kinematic viscosity (laminar diffusion) is set to 0 m$^2$/s.

$$F(r) = \int F(r')W(r - r', h)dr' \tag{8}$$

**Table 5**
Biochemical parameter values for lab-scale tank simulation.

| Parameter | Value | Unit |
|---|---|---|
| $K_{I,h2,c4}$ | 1E−5 | kg·COD·m$^{-3}$ |
| $K_{I,h2,fa}$ | 5E−6 | kg·COD·m$^{-3}$ |
| $K_{I,h2,pro}$ | 3.5E−6 | kg·COD·m$^{-3}$ |
| $K_{I,NH3}$ | 0.0018 | M |
| $K_{S,h2}$ | 7E−6 | kg·COD·m$^{-3}$ |
| $K_{S,IN}$ | 1E−4 | M |
| $K_{S,aa}$ | 0.3 | kg·COD·m$^{-3}$ |
| $K_{S,ac}$ | 0.15 | kg·COD·m$^{-3}$ |
| $K_{S,c4}$ | 0.2 | kg·COD·m$^{-3}$ |
| $K_{S,fa}$ | 0.4 | kg·COD·m$^{-3}$ |
| $K_{S,pro}$ | 0.1 | kg·COD·m$^{-3}$ |
| $K_{S,su}$ | 0.5 | kg·COD·m$^{-3}$ |
| $k_{hyd,ch}$ | 10 | d$^{-1}$ |
| $k_{hyd,li}$ | 10 | d$^{-1}$ |
| $k_{hyd,pr}$ | 10 | d$^{-1}$ |
| $k_{dec,Xh2}$ | 0.02 | d$^{-1}$ |
| $k_{dec,Xaa}$ | 0.02 | d$^{-1}$ |
| $k_{dec,Xac}$ | 0.02 | d$^{-1}$ |
| $k_{dec,Xc4}$ | 0.02 | d$^{-1}$ |
| $k_{dec,Xfa}$ | 0.02 | d$^{-1}$ |
| $k_{dec,Xpro}$ | 0.02 | d$^{-1}$ |
| $k_{dec,Xsu}$ | 0.02 | d$^{-1}$ |
| $k_{dis}$ | 0.5 | d$^{-1}$ |
| $k_{m,h2}$ | 35 | d$^{-1}$ |
| $k_{m,aa}$ | 50 | d$^{-1}$ |
| $k_{m,ac}$ | 8 | d$^{-1}$ |
| $k_{m,c4}$ | 20 | d$^{-1}$ |
| $k_{m,fa}$ | 6 | d$^{-1}$ |
| $k_{m,pro}$ | 13 | d$^{-1}$ |
| $k_{m,su}$ | 30 | d$^{-1}$ |
| $pH_{LL,h2}$ | 5 | – |
| $pH_{LL,aa}$ | 4 | – |
| $pH_{LL,ac}$ | 6 | – |
| $pH_{UL,h2}$ | 6 | – |
| $pH_{UL,aa}$ | 5.5 | – |
| $pH_{UL,ac}$ | 7 | – |

$$F(r_a) \approx \sum_b \frac{m_b}{\rho_b} F(r_b) W(r_a - r_b, h) \quad (9)$$

Applying this scheme to mass and momentum conservation of the continuum, the governing equation can be written as Eq. (10), where $\nu_0$ is the kinematic viscosity, $\boldsymbol{u}_a$, $P_a$ and $\rho_a$ are the velocity, pressure and density of particle $a$. And $\eta$ is a small number to ensure the denominators stay non-zero. The term $g$ denotes a source term, gravity for example.

$$\frac{d\boldsymbol{u}_a}{dt} = -\sum_b m_b \left(\frac{P_a + P_b}{\rho_a \rho_b}\right) \nabla_a W_{ab} + \sum_b m_b \left(\frac{4\nu_0}{\rho_a + \rho_b}\right)\left(\frac{r_{ab}}{r_{ab}^2 + \eta^2}\right) \nabla_a W_{ab} + \Gamma + g \quad (10)$$

Unlike Eulerian methods, the Lagrangian nature of SPH eliminates the need for the Reynolds Transport Theorem (RTT) in the advection term. This is because mass and momentum transport are inherently carried by the movement of particles. However, this can lead to an underestimation of the dissipative term, namely numerical dissipation, in the governing equations. To overcome this issue, artificial viscosity (Monaghan, 1992) and Sub-Particle Scale (SPS) (Lo and Shao, 2002), are introduced as an additional dissipative term Γ. In this work, only the SPS method is considered, and it will be discussed in the next section.

### 2.3. Turbulence modeling

The Sub-Particle Scale (SPS) turbulence modeling takes inspiration from the classic Large-Eddy Simulation (LES) turbulence modeling in the Eulerian method, where it assumes that the unresolved turbulence and turbulent kinetic energy (TKE) $k_{SPS}$ ($k_{SGS}$ for OpenFOAM implementation where SGS denoting Sub-Grid Scale), are modeled by an additional viscous stress tensor $\tau_{ij}$ and turbulent viscosity (eddy viscosity) $\nu_t$ and are calculated by Eqs. (12) and (14) (Dalrymple and Rogers, 2006).

$$\Gamma = \sum_b m_b \left(\frac{\boldsymbol{\tau}_{ij}^a}{\rho_a^2} + \frac{\boldsymbol{\tau}_{ij}^b}{\rho_b^2}\right) \nabla_a W_{ab} \quad (11)$$

$$\boldsymbol{\tau}_{ij} = \rho \nu_t \left(2\widetilde{S}_{ij} - \frac{2}{3}\widetilde{S}_{kk}\delta_{ij}\right) - \frac{2}{3}\rho C_I \Delta^2 \delta_{ij} \quad (12)$$

$$k_{SPS} = \frac{1}{2}(\overline{\boldsymbol{u}_k \boldsymbol{u}_k} - \overline{\boldsymbol{u}_k} \bullet \overline{\boldsymbol{u}_k}) \text{ or } \frac{1}{2}\boldsymbol{\tau}_{kk} \quad (13)$$

where $\delta_{ij}$ is the Kronecker delta, and $\overline{(\bullet)}$ denotes an arbitrary spatial filter. A spatial filter in the context of turbulence modeling describes a spatial resolution chosen by the Large-Eddy Simulation (LES) models. Further detailed information of the flow field (velocity fields, pressure field, temperature field, etc.) are "filtered" or averaged. In the standard Smagorinsky model, it indicates the implicit "filter" resulting from the finite resolution when discretizing the computational domain (Eulerian) or continuum (SPH). And $\widetilde{(\bullet)}$ denotes a Favre-averaging operation (averaging according to the density field) to incorporate the weakly-compressibility of SPH method, and $\widetilde{S}_{ij}$ is the Favre-averaged strain rate tensor. $C_I$ is an empirical constant with a value of 0.0066 (Blin et al., 2003). The eddy viscosity $\nu_T$ is calculated by the following equations, cell. For the SPH method, the radius of the smooth kernel is $2h$. And Eq. (15) can be rewritten as Eq. (16),

$$\left|\widetilde{S}_{ij}\right|^2 = 2\left(\widetilde{S}_{11}^2 + \widetilde{S}_{22}^2 + \widetilde{S}_{33}^2\right) + 4\left(\widetilde{S}_{12}^2 + \widetilde{S}_{13}^2 + \widetilde{S}_{23}^2\right) \quad (16)$$

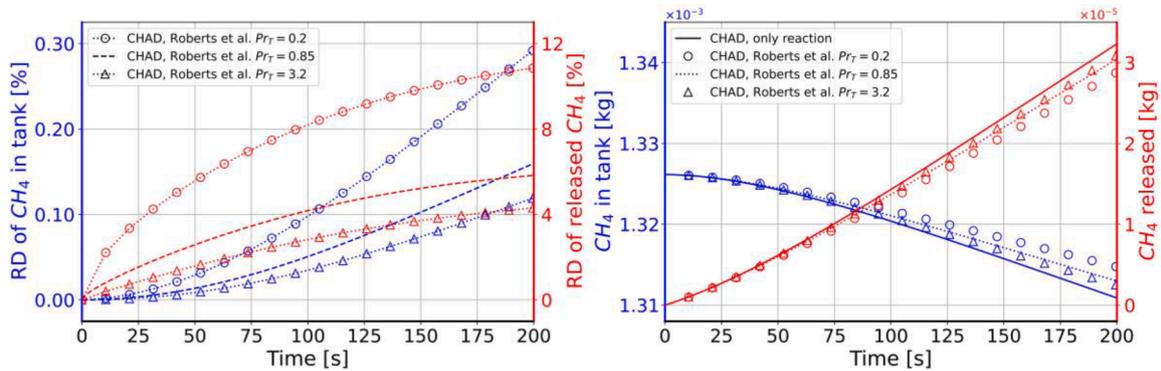

**Fig. 6.c.** The comparison of methane production of lab-scale reactor over 200 s with different turbulent Prandtl number for different models. Relative difference shown on the left and absolute mass shown in the right. All simulations are set to exclude the chemical turbulent diffusion (Sc$_T$ = 0). The kinematic viscosity (laminar diffusion) is set to be 1E−5 m$^2$/s.

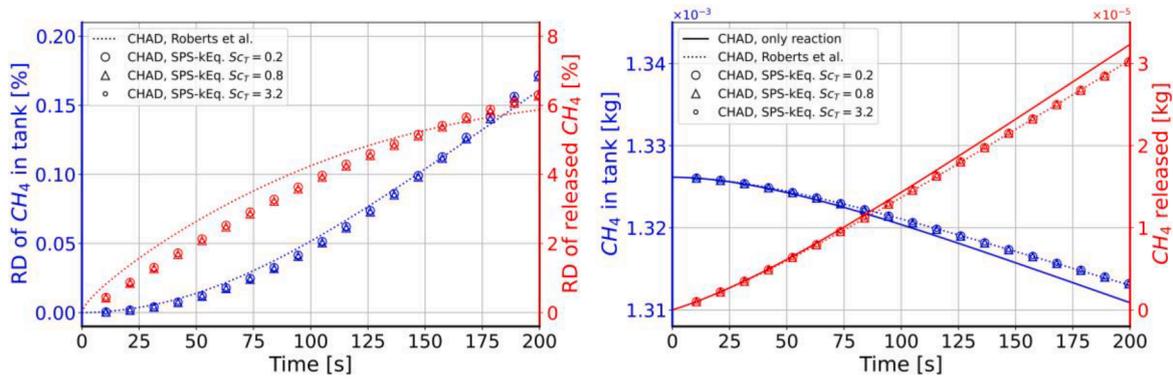

**Fig. 6.d.** The comparison of methane production of lab-scale reactor over 200 s with different turbulent Schmidt number for different models. All simulations are set to include all thermal and chemical diffusion.

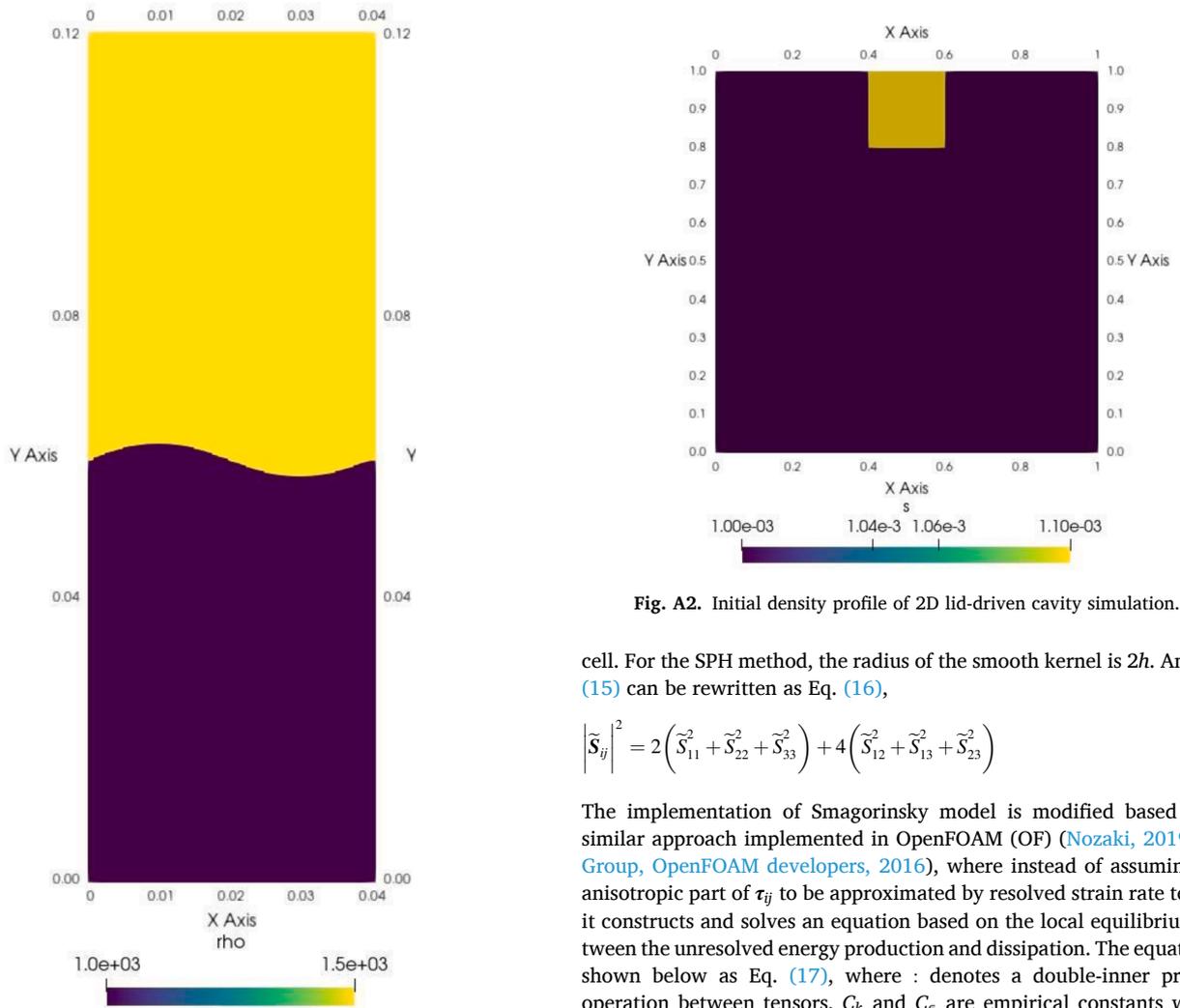

**Fig. A1.** Initial density profile of 2D Rayleigh-Taylor Instability simulation.

**Fig. A2.** Initial density profile of 2D lid-driven cavity simulation.

The implementation of Smagorinsky model is modified based on a similar approach implemented in OpenFOAM (OF) (Nozaki, 2019; ESI Group, OpenFOAM developers, 2016), where instead of assuming the anisotropic part of $\tau_{ij}$ to be approximated by resolved strain rate tensor, it constructs and solves an equation based on the local equilibrium between the unresolved energy production and dissipation. The equation is shown below as Eq. (17), where : denotes a double-inner product operation between tensors. $C_k$ and $C_\epsilon$ are empirical constants with a value of 0.094 and 1.048 respectively.

$$\nu_T = (C_s \Delta)^2 \left|\widetilde{S}_{ij}\right| \quad (14)$$

$$\left|\widetilde{S}_{ij}\right| = \left(2\widetilde{S}_{ij}\widetilde{S}_{ij}\right)^{\frac{1}{2}} \quad (15)$$

where $C_s$ is an empirical constant which varies depending on the application. $\Delta$ is the size of the filter. In the standard Smagorinsky model (Smagorinsky, 1963) the filter is implicitly chosen to be the size of the

$$\widetilde{S}_{ij} : \tau_{ij} + C_\epsilon \frac{k_{SPS}^{1.5}}{\Delta} = 0 \quad (17)$$

$$k_{SPS} = \frac{-b \pm \sqrt{b^2 - 4ac}}{2a} \quad (18)$$

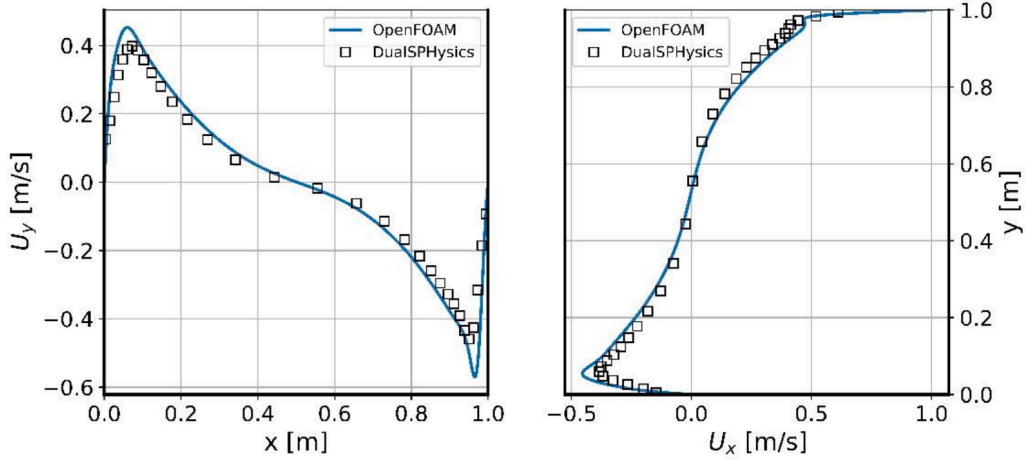

**Fig. A3.** Velocity profile of 2D lid-driven cavity simulation plotted along the center of the domain on y and x-axis.

$$\begin{cases} a = \dfrac{C_\epsilon}{\Delta} \\ b = \dfrac{2}{3} tr\left(\widetilde{S}_{ij}\right) \\ c = 2C_k \Delta \left( \text{dev}\left(\widetilde{S}_{ij}\right) : \widetilde{S}_{ij} \right) \end{cases}$$

where $tr\left(\widetilde{S}_{ij}\right)$ is the trace of tensor $\widetilde{S}_{ij}$, and $\text{dev}\left(\widetilde{S}_{ij}\right)$ is the deviatoric part of the tensor which is calculated as below in Eq. (19). With the solution for $k_{SPS}$, $\nu_T$ can be found using Eq. (20).

$$\text{dev}\left(\widetilde{S}_{ij}\right) = \widetilde{S}_{ij} - \frac{1}{3} tr\left(\widetilde{S}_{ij}\right) \tag{19}$$

$$\nu_T = C_k \Delta \sqrt{k_{SPS}} \tag{20}$$

### 2.4. Turbulence diffusion

The diffusion of scalars in the flow fields is defined by Eq. (21) as follows. $\varphi$ is the diffused scalar. $D_{\text{eff}}$, also known as the effective diffusion coefficient in some sources, represents the effective diffusivity of the scalar with a dimension of $(kg \bullet m^{-1} \bullet s^{-1})$. In the later section, the diffusivity will be expressed differently depending on whether it is for chemical concentrations or temperature of the fluid. Eq. (21) represents a general form for scalar diffusion where $d$ is used to indicate the time derivative on the left-hand side (LHS), differentiate from $\partial$ used for the special derivative on the right-hand side (RHS). This distinction arises from whether the equation is formed from a Eulerian or Lagrangian perspective. As mentioned previously in Eulerian methods the conservation laws are applied to a fixed mesh, necessitating the use of the Reynolds Transport Theorem to account for advection between adjacent mesh cells. In Lagrangian method, SPH namely, the advection of the scalars is modeled naturally by the movement of the particles, eliminating the need for additional terms.

$$\frac{d\varphi}{dt} = \frac{D_{\text{eff}}}{\rho} \frac{\partial^2 \varphi}{\partial x_i^2} \tag{21}$$

Eq. (21) above is rewritten in to Eqs. (22) and (23) for chemical concentrations and temperature of the fluid for SPH convention (Monaghan et al., 2005), where $C_{p,a}$ is the specific thermal capacity of the particle $a$ with the dimension of $(J \bullet kg^{-1} \bullet K^{-1})$, and $\kappa_a$ is the thermal conductivity of the particle $a$ with the dimension of $(W \bullet m^{-1} \bullet K^{-1})$.

$$\frac{dC_a}{dt} = \sum_b \left(\frac{m_b}{\rho_a \rho_b}\right) \left(\frac{4D_{\text{eff},a} D_{\text{eff},b}}{D_{\text{eff},a} + D_{\text{eff},b}}\right) (C_a - C_b) F_{ab} \tag{22}$$

$$C_{p,a} \frac{dT_a}{dt} = \sum_b \left(\frac{m_b}{\rho_a \rho_b}\right) \left(\frac{4\kappa_{\text{eff},a} \kappa_{\text{eff},b}}{\kappa_{\text{eff},a} + \kappa_{\text{eff},b}}\right) (T_a - T_b) F_{ab} \tag{23}$$

$$F_{ab} = \left(\frac{r_{ab}}{r_{ab}^2 + \eta^2}\right) \nabla_a W_{ab}$$

In this study, the specific thermal capacity of the particles is considered to be constant for all fluid particles of the same type of fluid. For the sake of simplicity, the thermal diffusion equation above and the thermal diffusivity can be rewritten in Eqs. (24) and (25) as follows. $\alpha_{\text{eff}}$ is the effective thermal diffusivity and shares the same dimension with $D_{\text{eff}}$.

$$\frac{dT_a}{dt} = \sum_b \left(\frac{m_b}{\rho_a \rho_b}\right) \left(\frac{4\alpha_{\text{eff,a}} \alpha_{\text{eff,b}}}{\alpha_{\text{eff,a}} + \alpha_{\text{eff,b}}}\right) (T_a - T_b) F_{ab} \tag{24}$$

$$\alpha_{\text{eff}} = \frac{\kappa_{\text{eff}}}{C_p} \tag{25}$$

As shown in Eqs. (26) and (27) below, the effective diffusivity can be modeled as the sum of laminar diffusivity (neglecting molecular diffusivity) and turbulent diffusivity, denoted by subscript T, using an analogy to the SPS turbulence modeling discussed in the previous chapter. One of the commonly used methods to close for the turbulent diffusivity is the Gradient-Diffusion Hypothesis (GDH), which assumes isotropic turbulence and a linear relationship between the unresolved transportation of the mass and the passive scalar (Combest et al., 2011). Non-dimensional numbers like Prandtl number (*Pr*) and Schmidt number (*Sc*) are used, representing the linearity of the said relationship, shown in Eqs. (28) and (29), where $\mu$ is the dynamic viscosity with a dimension of $(kg \bullet m^{-1} \bullet s^{-1})$ and $\nu$ is the kinematic viscosity with a dimension of $(m^2 \bullet s^{-1})$. For choosing the appropriate Prandtl and Schmidt number, there are no universal solutions. It depends on the specific flow type and application (Tominaga and Stathopoulos, 2007; Gualtieri et al., 2017).

$$\alpha_{\text{eff}} = \alpha + \alpha_T \tag{26}$$

$$D_{\text{eff}} = D + D_T \tag{27}$$

$$Pr = \frac{\mu}{\alpha} = \frac{\nu}{\frac{\alpha}{\rho}}, Pr_T = \frac{\mu_T}{\alpha_T} = \frac{\nu_T}{\frac{\alpha_T}{\rho}} \tag{28}$$

$$Sc = \frac{\mu}{D} = \frac{\nu}{\frac{D}{\rho}}, Sc_T = \frac{\mu_T}{D_T} = \frac{\nu_T}{\frac{D_T}{\rho}} \tag{29}$$

$$D_{\text{eff}} = \rho(\frac{\nu}{Sc} + \frac{\nu_{\text{T}}}{Sc_{\text{T}}})$$

$$\alpha_{\text{eff}} = \rho(\frac{\nu}{Pr} + \frac{\nu_{\text{T}}}{Pr_{\text{T}}})$$

Roberts and Webster (2003) proposed an alternative method where turbulent diffusivity $D_{\text{T}}$ is closed by the characteristic length and velocity of the diffused particle, shown in Eq. (30). The characteristic length $l_a$ for diffusion of particle $a$ is the approximation of the distance that the passive scalar can travel within the particle travel time. In the context of SPH scheme, it is chosen to be the kernel size $2h$. The characteristic velocity $\widetilde{v}_a$ for diffusion of particle $a$ is the variance of the velocity fluctuation $v'_a$, shown in Eq. (31). In this work, the variance is weighted using the smoothing kernel to account for the SPH scheme, expressed as Eq. (32).

$$D_{\text{T},a} = \rho_a \widetilde{v}_a l_a \tag{30}$$

$$v'_a = v_a - \sum_b \frac{m_b}{\rho_b} v_a W_{ab} \tag{31}$$

$$\widetilde{v}_a^2 = \overline{v'_a v'_a} = \frac{1}{N_b} \sum_b \frac{m_b}{\rho_b} (v'_a v'_a) W_{ab} \tag{32}$$

$$D_{\text{eff}} = \frac{\rho \nu}{Sc} + D_{\text{T}}$$

$$\alpha_{\text{eff}} = \frac{\rho \nu}{Pr} + \frac{D_{\text{T}} Sc_{\text{T}}}{Pr_{\text{T}}}$$

Another approach to modeling turbulent diffusivity was proposed by Greif et al. (2009), who directly modeled the effective diffusivity $D_{\text{eff}}$ rather than decomposing it into laminar and turbulent components. Similar to the method of Roberts et al. (Rosén and Jeppsson, 2006), the closure utilizes characteristic length and velocity of the diffused particle. The characteristic length is the same as in Roberts et al. ($2h$), while the characteristic velocity is chosen as the velocity dispersion, as shown in Eq. (34), where subscript $a$ indicates the local diffused particle and $N_b$ is the number of the neighboring particles within the smoothing kernel. $v_a$ and $v_b$ are the bulk velocity of particle $a$ and $b$ respectively.

$$D_{\text{eff},a} = \rho_a \widetilde{v}_a l_a \tag{33}$$

$$\widetilde{v}_a^2 = \frac{1}{N_b} \sum_b |v_a - v_b|^2 \tag{34}$$

Although Combest et al. (2011) mentioned other methods, such as Algebraic models and scalar-flux models, for modeling the turbulent diffusivity, these methods were not investigated in this study due to the complexity and computational cost involved in solving the additional transport equations for TKE and turbulence dissipation rate. Future studies may explore these models to provide a more comprehensive understanding of turbulent diffusivity in SPH simulations. In this work, the models proposed by Roberts et al. and Greif et al. would be investigated together with the nondimensional number approach (SPS model), and the results are shown in the later chapter.

## 3. Case studies

The implementation of the turbulent diffusion models is validated with a gravity-driven 2D Rayleigh-Taylor Instability and a 2D lid driven cavity case. After the validation, a real-life based lab-scale anaerobic digestion tank (Neuner et al., 2022) is tested with CHAD simulation to verify the effectiveness of the turbulent diffusion modeling. The initial states of all the cases are graphically presented in the appendix.

### 3.1. Rayleigh-Taylor Instability

In this case, a 2D Rayleigh-Taylor Instability (RTI) simulation is conducted, where two fluids with different densities and temperatures are considered. The simulation is gravity-driven, with the force of gravity acting in the $-y$ direction. The obtained results are compared with the outcomes from the OpenFOAM simulations. The simulation parameters are provided in Table 1. To focus on the effect of turbulent diffusion, the thermal conductivity is deliberately kept small to minimize the influence of resolved thermal diffusion. Additionally, a small turbulent Prandtl number is chosen to enhance the impact of turbulent diffusion.

### 3.2. Lid-driven cavity

To validate the implementation of turbulent diffusion models, a 2D lid-driven cavity case is employed. The parameters for this study case are listed in Table 2. In this scenario, instead of examining thermal properties, a passive scalar tracer "s" is used to investigate the diffusion characteristics of the flow field. Similar to the previous case, the laminar part of diffusion is neglected by setting a high Schmidt number. A turbulent Schmidt number of 0.1 is selected to emphasize the influence of turbulent diffusion.

### 3.3. Lab-scale AD tank

Another test case is based on the geometry of a real-world lab-scale digester (Neuner et al., 2022). The simulation is initialized with two layers of sludge, each having different initial temperatures of 308.15 K and 309.15 K, along with different concentrations of chemical components following the ADM1 model. The reactor is designed to operate under thermal conditions appropriate for anaerobic digestion reactions. It is configured as a batch reactor, representing a closed system in terms of mass exchange. A helical mixer is employed to stir the fluid at a constant rotational speed, while the reactor walls and mixer have distinct constant temperatures. All the relevant parameters for this case are listed in Table 3 below. The initial condition of the tank is presented in Fig. 1.a..

## 4. Results and discussion

### 4.1. Rayleigh-Taylor Instability

The results of 2D RTI simulations of OF and CHAD are presented in Fig. 2.a. and 2.b.. Despite the best efforts spent tweaking the parameters to have comparable flow development, Eulerian method and Lagrangian method are too different handling multiphase simulations. The temperature and density profile are averaged along x-axis and plotted along y-axis. As can be seen from Fig. 2.a., the temperature profile has a relatively good agreement between OF and CHAD results in the earlier stage of the dynamic. However, the profile diverges gradually as time progresses. To validate how much of the divergence is stemming from the disagreement from the interface between the 2 layers of fluid, in Fig. 2.b. the relative difference between average density and temperature profile along y-axis are compared. It is clear from Fig. 2.b. that most of the disagreement in temperature between OF and CHAD simulation is resulting from the difference in density. In Fig. 5 the differences of temperature profile with and without turbulent diffusion in OF and CHAD are compared. As the movement of 2 layers of fluid diverges and the interface between the two develop differently, the profile of the 2 cases do not follow each other exactly, but for the bulk part within the same range. To validate the implementation of the turbulent diffusion models with a better matching flow development, another validation case based on 2D lid-driven cavity simulation is created and the results are shown in section 4.2.

*4.2. Lid-driven cavity*

The results of 2D lid-driven cavity cases are shown in Figs. 3–5. The velocity field is comparable between OF and CHAD simulation once the flow fields are fully developed. The velocity profile of x and y-axis plotted against y and x-axis respectively, and with these flow fields the implementation of the models can be validated better than the 2D RTI cases, shown by figures in Appendix.

Figs. 3.a. and 3.b. show the comparison of the concentration fields of and chad simulation without diffusion at $t = 2s$ and $t = 4s$. The discrete separation of concentrations of CHAD simulations is visible. With a comparable velocity field between OF and CHAD simulations, a better agreement of concentration profile development, shown in Fig. 4.a. and 4.b., between OF and CHAD can be observed. Figs. 5.a. and 5.b. show the concentration profiles of the cases with 2 different turbulent diffusion models, plotted and compared along x and y-axis shown by the redlines in Figs. 4.a. and 4.b. Visually the results of CHAD and OF agrees quite nicely.

*4.3. Lab-scale reactor*

After the implementations being validated, a lab-scale AD reactor tank based on a real-life model (Neuner et al., 2022) is tested. The study case is simulated in 3D the results are presented as below. Fig. 1.b. shows the concentration fields of soluble and gaseous methane and temperature field after 200 s of simulation, with both thermal and chemical turbulent diffusion model implemented. The inhomogeneity within the tank is obvious, even with an active mixing device. A relatively poorly mixed region can be spotted at the bottom of the tank, and the region near the mixing device has better mixing due to the turbulence created by the mixer, highlighting the importance of including CFD and turbulent diffusion models into such simulations.

In Fig. 6.a., the results in terms of methane production with different turbulent diffusion models and no thermal turbulent diffusion are compared. On the left is the relative differences (RD) of CHAD between the results with only biochemical, and the results with turbulent diffusion model proposed by Greif and Roberts. In Fig. 6, the blue line and marks are the RD of the total amount of methane (soluble plus gaseous) within the tank while the red show the RD of the amount of gaseous methane released from the tank due to pressure balance with the atmosphere pressure. The plot on the right is the mass of methane stored within the tank (blue) and release as gas (red). With kinematic viscosity set to be $1E-5\ m^2/s$, the model proposed by Roberts and Webster (2003) gives comparable results as the model proposed by Greif et al. (2009). As shown by the plot on the right of Fig. 6.a., the difference between the results with only reaction and with diffusion is noticeable, while the different values of kinematic viscosity leave trivial impact on the results.

To compare the impact of the turbulent Schmidt number $Sc_T$, the simulations are set to have no kinematic viscosity and no thermal diffusion. Since the model proposed by Roberts and Webster (2003) is independent of $Sc_T$, it is selected to be the calibration for finding the best $Sc_T$ value for this case. It is found that when $Sc_T = 0.2$ the results of SPS turbulent diffusion model are comparable with Roberts and Webster (2003) as indicated by Fig. 6.b., which is aligned with the values proposed by Gualtieri et al. (2017). From the RD plot of Fig. 6.b., it can be seen that with higher $Sc_T$, meaning less chemical turbulent diffusion, the result is closer to the result where only biochemical reaction is applied, showing the effect of chemical mixing on the efficiency of the reaction tank (see Table 5).

For the following simulations, different choices of turbulent Prandtl number are tested and the results are shown in Fig. 6.c.. The effect chemical turbulent diffusion is excluded by setting $D_{eff} = D$. From the finding shown in Fig. 6.a., the kinematic viscosity (laminar diffusion) is set to be $1E-5\ m^2/s$. As can be observed, the thermal turbulent diffusion has much higher impact on the biochemical reactions within the reaction tank compared to the chemical turbulent diffusion. With higher $Pr_T$, meaning less thermal turbulent diffusion, the result diverges less compared to the result where only biochemical reaction is applied, showing the effect of thermal mixing on the efficiency of the reaction tank.

With the findings from Figs. 6.a., 6.b. and 6.c. the kinematic viscosity is set to be 1E-5 $m^2/s$ and turbulent Prandtl number $Pr_T$ is set to be 0.85 for the other simulations. The simulations are made for Roberts model and SPS model with different turbulent Schmidt number $Sc_T$. The results are shown in Fig. 6.d.. Now with both thermal and chemical turbulent diffusion implemented, the 2 turbulent diffusion models show different impact on the reaction, as shown by the RD plot of Fig. 6.d. However, no meaningful difference between the models is spotted on the absolute mass of the methane production, which is expected since chemical turbulent diffusion has less impact on the reaction compared to thermal turbulent diffusion and all simulations have the same $Pr_T$.

*4.4. Computational cost*

The pre-processing of data (computational fluid dynamic simulations) is one limitation that cannot be overlooked, however it is not within the scope of this study. A thorough study conducted by Kumar et al. (2022) had covered this topic. In the current implementation of CHAD, simulating the 3D lab-scale reaction tank in the study case on an AMD Ryzen 7 5800X CPU (parallelized among 12 computational threads), excluding the thermal and chemical diffusion, takes around 2800 s for a 200-second reaction. With the constant thermal and chemical diffusion included, the same simulation is roughly 75 % slower, finishing around 4900 s. Adding the turbulent diffusion models results in an extra 0.3–0.5 % time increase.

## 5. Conclusion

Two turbulent diffusion models implemented and validated for CHAD produce comparable results, with the best agreement achieved at $Sc_T$ set to 0.2 for this specific setup. Within a 200-second batch reactor, mixing has minimal impact on biochemical reactions, while thermal turbulent diffusion exerts a greater influence than chemical turbulent diffusion. These findings highlight the effectiveness of turbulent diffusion models for coupling CFD with biochemical reactions, especially in scenarios with rapid dynamics. As reaction dynamics accelerate, turbulent diffusion models become increasingly crucial. Further investigations are needed for scenarios involving multiphase flows.

**CRediT authorship contribution statement**

**Jeremy Z. Yan:** . **Prashant Kumar:** Conceptualization, Methodology, Software, Writing – review & editing. **Wolfgang Rauch:** Conceptualization, Funding acquisition, Methodology, Project administration, Resources, Supervision, Writing – review & editing.

**Declaration of Competing Interest**

The authors declare the following financial interests/personal relationships which may be considered as potential competing interests: Jeremy Yan reports financial support was provided by European Commission.

**Data availability**

Data will be made available on request.

**Acknowledgments**

This work is supported by the funding from the European Union's Horizon 2020 research and innovation program under the Marie

Skłodowska-Curie grant agreement No. 847476. The contents of this publication do not necessarily reflect the position or opinion of the European Commission.

**Appendix**

See Figs. A1–A3